\documentclass[12pt]{article}
\usepackage{amsmath,amsthm}
\usepackage{txfonts,bm,pifont}  
\usepackage{cite}  
\usepackage{graphicx,xcolor}
\usepackage[dvipdfm,bookmarks=true,bookmarksnumbered=true,colorlinks=true]{hyperref}
\newtheorem{theorem}{Theorem}[section]
\newtheorem{proposition}[theorem]{Proposition}

\newtheorem{remark}[theorem]{Remark}

\def\Z{{\mathbb{Z}}} 
\makeatletter
\@addtoreset{equation}{section}
\renewcommand{\theequation}{\@arabic\c@section.\@arabic\c@equation}
\long\def\@makecaption#1#2{
 \vskip 10pt 
 \setbox\@tempboxa\hbox{#1. #2}
 \ifdim \wd\@tempboxa >\hsize #1. #2\par \else \hbox
to\hsize{\hfil\box\@tempboxa\hfil} 
 \fi}
\makeatother
\begin{document}
\begin{center}
\renewcommand{\baselinestretch}{1.3}\selectfont
\begin{large}
\textbf{Bilinearization and Special Solutions to the Discrete Schwarzian KdV Equation}
\end{large}\\[4mm]
\renewcommand{\baselinestretch}{1}\selectfont
\textrm{\large Mike Hay$^1$, Kenji Kajiwara$^2$,
and Tetsu Masuda$^3$}\\[2mm]
$^1$: Faculty of Mathematics, Kyushu University, 744 Motooka, Fukuoka 819-8581, Japan.\\
hay@math.kyushu-u.ac.jp
\\
$^2$: Faculty of Mathematics, Kyushu University, 744 Motooka, Fukuoka 819-8581, Japan.\\
kaji@math.kyushu-u.ac.jp
\\
 $^2$: Department of Physics and Mathematics, Aoyama Gakuin University, Sagamihara 229-8558, Japan.
\\
masuda@gem.aoyama.ac.jp
\\[2mm]
28 February 2011
\end{center}
\begin{abstract}
Various solutions to the discrete Schwarzian KdV equation are discussed.  We first derive the
bilinear difference equations of Hirota type of the discrete Schwarzian KP equation, which is
decomposed into three discrete two-dimensional Toda lattice equations. We then construct two kinds
of solutions in terms of the Casorati determinant.  We derive the discrete Schwarzian KdV equation
on an inhomogeneous lattice and its solutions by a reduction process. We finally discuss the
solutions in terms of the $\tau$ functions of some Painlev\'e systems.
\end{abstract}



\section{Introduction}
The focus of this article is on the partial difference equation
\begin{align}
 \frac{(z_{l_1,l_2}-z_{l_1+1,l_2})(z_{l_1+1,l_2+1}-z_{l_1,l_2+1})}{(z_{l_1+1,l_2}-z_{l_1+1,l_2+1})(z_{l_1,l_2+1}-z_{l_1,l_2})}
=\frac{\lambda(l_1)}{\mu(l_2)},\label{eqn:cross_ratio}
\end{align}
known as the {\it discrete Schwarzian KdV equation} (dSKdV), where $l_k$ ($k=1,2$) are independent
variables, $z_{l_1,l_2}$ denotes the value of the dependent variable $z$ at the lattice site
$(l_1,l_2)$ and $\lambda(l_1)$ and $\mu(l_2)$ are arbitrary functions in the indicated variables. As
a lattice equation, dSKdV (\ref{eqn:cross_ratio}) was first studied in
\cite{NQC83,Nijhoff-Capel:dKdV}, and classified as a special case of `Q1' equation in the ABS list
\cite{ABS}. It yields the Schwarzian KdV equation 
\begin{equation}
\psi_t = \psi_xS(\psi),\quad S(\psi)\equiv \frac{\psi_{xxx}}{\psi_x} - \frac{3}{2}\frac{\psi_{xx}^2}{\psi_x^2},\label{eqn:SKdV}
\end{equation}
in the continuous limit, and is related to the lattice modified KdV equation and the lattice KdV
equation by Miura transformations \cite{NRGO:P6}. Note that the differential operator $S$ in
(\ref{eqn:SKdV}) is the {\it Schwarzian derivative} and thus (\ref{eqn:SKdV}) is invariant with
respect to M\"obius transformations. The soliton equations with M\"obius invariance 
may consequently be called `Schwarzian'. See \cite{n96} for a review of Schwarzian equations. We
sometimes refer to (\ref{eqn:cross_ratio}) as {\it non-autonomous} for non-constant $\lambda(l_1)$
and $\mu(l_2)$. When $\lambda(l_1)$ and $\mu(l_2)$ are both constants, we call it {\it
autonomous}. dSKdV is also known as the {\it cross-ratio equation}, since the left hand side of
(\ref{eqn:cross_ratio}) is the cross ratio, which is fundamental to many branches of geometry and
was studied as early as the time of Euclid. Our main motive for exploring this system comes from a
geometrical setting: dSKdV arises as one of the most basic equations in the discrete differential
geometry, which is expected to provide a new mathematical framework of discretization of
various geometrical objects. For example, dSKdV is used to define the discrete conformality of
discrete isothermic nets \cite{Bobenko-Pinkall:Discrete_Isothermic,Bobenko-Pinkall:Oxford}.

For notational simplicity, we display only the shifted
independent variables. For example,  $z_{l_1+1,l_2}$ may be written as $z_{l_1+1}$. Using this convention,
(\ref{eqn:cross_ratio}) may be written as
\begin{align}
 \frac{(z-z_{l_1+1})(z_{l_1+1,l_2+1}-z_{l_2+1})}{(z_{l_1+1}-z_{l_1+1,l_2+1})(z_{l_2+1}-z)}
=\frac{\lambda(l_1)}{\mu(l_2)}.\label{eqn:cross_ratio2}
\end{align}

Note that if there exists a function $u_{l_1,l_2}$
satisfying
\begin{equation}
 z_{l_i+1}-z = f_i(l_i)~u_{l_i+1}u \quad (i=1,2)\label{eqn:red_bilinear}
\end{equation}
where $f_i(l_i)$ ($i=1,2$) are arbitrary functions, then (\ref{eqn:cross_ratio}) is
automatically satisfied with $\lambda(l_1)=f_1(l_1)^2$ and $\mu(l_2)=f_2(l_2)^2$.

In this article, we aim to clarify the structure of the bilinear difference equations and $\tau$
functions associated with dSKdV (\ref{eqn:cross_ratio}).  We then construct various solutions to
dSKdV, including the soliton type solutions of autonomous dSKdV presented in
\cite{HZ:ABS2,NAH:ABS1}. In section 2, we consider the {\it discrete Schwarzian KP equation}
(dSKP). We first discuss the bilinear equations and $\tau$ function of dSKP in section
\ref{subsec:dSKP}. We then establish the reduction procedure to derive dSKdV from dSKP. In
\cite{Konopelchenko-Schief} a basic geometric property of circles was shown to be equivalent to
dSKP, and dSKdV as a special case. In that article a reduction from dSKP to a degenerate dSKdV was
discussed. The full, non-autonomous dSKdV is difficult to achieve via reduction because of
complications arising from the non-autonomous terms. However, a method to circumvent these
complications was described in \cite{KO:non-autonomous_dKdV,KO:non-autonomous:RIMS}, where the
reduction is performed along auxiliary variables, enabling the reduced equation to remain
non-autonomous in its independent variables. We use a similar technique to obtain dSKdV from dSKP by
reduction. We thereby construct solutions to (\ref{eqn:cross_ratio}) in the form of $\tau$ functions
of the Casorati ($N$-soliton) and molecule types. This is explained in section \ref{sec:dSKP_red}.

dSKdV appears not only as one of the discrete soliton equations, but also as the equation describing
the chain of B\"acklund transformations of the Painlev\'e systems. This was first reported in
\cite{NRGO:P6} for the case of the Painlev\'e VI equation, which implies that dSKdV also admits
solutions expressible in terms of the $\tau$ functions of the Painlev\'e VI equation. The same
situation also arises for other Painlev\'e systems. In section 3, we discuss dSKdV in the setting of
the Painlev\'e systems and construct explicit formulae of solutions in terms of their $\tau$
functions.  We consider the Painlev\'e systems with the affine Weyl group symmetry of type
$(A_2+A_1)^{(1)}$ and $D_4^{(1)}$, where the former includes a $q$-Painlev\'e III equation, and the
latter the Painlev\'e VI equation.
\section{Discrete Schwarzian KP Equation and Its Reduction}\label{sec:dSKP}
\subsection{Discrete Schwarzian KP Equation}\label{subsec:dSKP}
The discrete Schwarzian KP equation (dSKP) is 
\begin{equation}
\frac{(z_{l_1+1}-z_{l_1+1,l_2+1})(z_{l_2+1}-z_{l_2+1,l_3+1})(z_{l_3+1}-z_{l_1+1,l_3+1})}
{(z_{l_1+1,l_2+1}-z_{l_2+1})(z_{l_2+1,l_3+1}-z_{l_3+1})(z_{l_1+1,l_3+1}-z_{l_1+1})}= -1 \label{eqn:dSKP}
\end{equation}
where $l_i$ $(i=1,2,3)$ are the discrete independent variables and $z=z_{l_1,l_2,l_3}$ is the
dependent variable. dSKP was first published in an alternative form in \cite{ncwq84}, first appeared
in the quoted form in \cite{dn91} and was also studied by \cite{Konopelchenko-Schief}. In the context
of discrete differential geometry, $z$ arises as a complex valued function. We first give an
explicit formula for the solution of dSKP in terms of a $\tau$ function:
\begin{proposition}\label{prop:D2DTL}
Let $\tau_{l_1,l_2,l_3}^{m,s}$ be the $\tau$ function satisfying the following bilinear
difference equations:
\begin{align}
\tau_{l_i+1}^{m+1}\tau
- \tau_{l_i+1}\tau^{m+1}
 =\tau_{l_i+1}^{s+1}\tau^{m+1,s-1}\quad (i=1,2,3), \label{bl:D2DTL}
\end{align}
 where $m$ and $s$ are the auxiliary independent variables. Then,
\begin{equation}
z_{l_1,l_2,l_3} = \frac{\tau_{l_1,l_2,l_3}^{m+1,s}}{\tau_{l_1,l_2,l_3}^{m,s}},\label{dSKP:z}
\end{equation}
satisfies dSKP (\ref{eqn:dSKP}).
\end{proposition}
\begin{proof}
First note that if there exist some functions $v=v_{l_1,l_2,l_3}$ and
$w=w_{l_1,l_2,l_3}$ such that
\begin{equation}
 z_{l_i+1}-z = v_{l_i+1}w\quad (i=1,2,3),\label{dSKP_ansatz}
\end{equation}
then (\ref{eqn:dSKP}) is automatically satisfied. Dividing (\ref{bl:D2DTL}) by $\tau_{l_i+1}\tau$ we have
\begin{displaymath}
\frac{\tau_{l_i+1}^{m+1}}{\tau_{l_i+1}}
- \frac{\tau^{m+1}}{\tau}
 =\frac{\tau_{l_i+1}^{s+1}}{\tau_{l_i+1}}\frac{\tau^{m+1,s-1}}{\tau}\quad (i=1,2,3), 
\end{displaymath}
which is equivalent to (\ref{dSKP_ansatz}) if we define $v$ and $w$ by
\begin{equation}
 v = \frac{\tau^{s+1}}{\tau},\quad w=\frac{\tau^{m+1,s-1}}{\tau},\label{dSKP:uv}
\end{equation}
respectively. 
 \end{proof}
\begin{remark}
dSKP (\ref{eqn:dSKP}) is invariant under the change of the independent
variables $l_i\to -l_i$ ($i=1,2,3$). Therefore if the $\tau$ function
is a solution to the bilinear equation
\begin{equation}
 \tau^{m+1}\tau_{l_i+1}
- \tau\tau_{l_i+1}^{m+1}
 =\tau^{s+1}\tau_{l_i+1}^{m+1,s-1}\quad (i=1,2,3), \label{bl2:D2DTL}
\end{equation}
then $z$ in (\ref{dSKP:z}) also satisfies (\ref{eqn:dSKP}). In this case, $v$ and $w$
in (\ref{dSKP_ansatz}) are expressed as
\begin{equation}
  v = \frac{\tau^{m+1,s-1}}{\tau},\quad w=-\frac{\tau^{s+1}}{\tau}.
\end{equation}
\end{remark}

Each of the bilinear equations in (\ref{bl:D2DTL}) is the {\it
discrete two-dimensional Toda lattice equation}\cite{Hirota:d2DTL} with respect to
the independent variables $(l_i,m,s)$ ($i=1,2,3$). It
is therefore possible to construct the Casorati determinant solution to these equations
(\ref{bl:D2DTL}) as follows \cite{Hirota:d2DTL,OKMS:RT,KS:q2DTL,NTSWK:non-autonomous2DTL}:
\begin{theorem}\label{thm:Casorati_tau:dSKP}
 Let $\sigma=\sigma_{l_1,l_2,l_3}^{m,s}$ be an $N\times N$ Casorati determinant
 defined by
\begin{equation}
 \sigma = \left|\begin{array}{cccc}
\varphi_1& \varphi_1^{s+1}&\cdots &\varphi_1^{s+N-1} \\
\varphi_2& \varphi_2^{s+1}&\cdots &\varphi_2^{s+N-1} \\
\vdots & \vdots &\cdots & \vdots\\
\varphi_N& \varphi_N^{s+1}&\cdots &\varphi_N^{s+N-1} 
\end{array}\right|,
\end{equation}
where $\varphi_i=\varphi_{i,l_1,l_2,l_3}^{m,s}$ ($i=1,\ldots,N$) are arbitrary functions satisfying
 the following linear relations:
\begin{align}
& \frac{\varphi_{i}-\varphi_{i,l_k-1}}{a_k(l_k-1)} = \varphi_{i}^{s+1}\quad (k=1,2,3),\label{dSKP:linear1}\\
& \frac{\varphi_{i}-\varphi_{i}^{m-1}}{b} = -\varphi_{i}^{s-1}.\label{dSKP:linear2}
\end{align}
Here, $a_k(l_k)$ ($k=1,2,3$) are arbitrary functions in the
 indicated independent variables and $b$ is a constant. Then 
\begin{equation}
 \tau =
  \left[\prod_{k=1}^3\prod_{j_k}^{l_k-1}(1+a_k(j_k)b)^{m}a_k(j_k)^s\right]b^{-ms}~\sigma \label{sig:D2DTL}
\end{equation}
satisfies the bilinear equations (\ref{bl:D2DTL}).
\end{theorem}
\begin{proof}
 It is known that $\sigma$ satisfies the bilinear equation
\begin{equation}
 (1+a_k(l_k)b)\sigma_{l_k+1}^{m+1}\sigma - \sigma_{l_k+1}\sigma^{m+1} 
 = a_k(l_k)b~\sigma_{l_k+1}^{s+1}\sigma^{m+1,s-1}\quad (k=1,2,3),\label{bl:sigma}
\end{equation}
which follows from an appropriate Pl\"ucker relation \cite{KS:q2DTL,NTSWK:non-autonomous2DTL,OHTI:dKP,OKMS:RT}. From
 (\ref{sig:D2DTL}) it is easily verified that $\tau$ satisfies (\ref{bl:D2DTL}).
\end{proof}
It is also possible to choose the size of the Casorati determinant as one of
the discrete independent variables in $l_k$ ($k=1,2,3$). This type of
solution is sometimes referred to as the {\it molecule type}.
\begin{theorem}\label{thm:dSKP_molecule}
 Let $\kappa=\kappa_{l_1,l_2,l_3}^{m,s}$ be an $l_1\times l_1$ Casorati determinant
 defined by
\begin{equation}
 \kappa = \left|\begin{array}{cccc}
\phi_1& \phi_1^{s+1}&\cdots &\phi_1^{s+l_1-1} \\
\phi_2& \phi_2^{s+1}&\cdots &\phi_2^{s+l_1-1} \\
\vdots & \vdots &\cdots & \vdots\\
\phi_{l_1}& \phi_{l_1}^{s+1}&\cdots &\phi_{l_1}^{s+l_1-1} 
\end{array}\right|,
\end{equation}
where $\phi_i=\phi_{i,l_2,l_3}^{m,s}$ ($i=1,\ldots,l_1$) satisfy
 the following linear relations:
\begin{align}
& \frac{\phi_{i,l_k+1}-\phi_{i}}{c_k(l_k)} = -\phi_{i}^{s+1}\quad (k=2,3),\label{dSKP:linear1_molecule}\\
& \frac{\phi_{i}-\phi_{i}^{m-1}}{b} = -\phi_{i}^{s-1}.\label{dSKP:linear2_molecule}
\end{align}
Here, $c_k(l_k)$ ($k=2,3$) are arbitrary functions in the
 indicated independent variables and $b$ is a constant. Then 
\begin{equation}
 \tau =
  \left[\prod_{k=2}^3\prod_{j_k}^{l_k-1}(1+c_k(j_k)b)^{-m}c_k(j_k)^{-s}\right]b^{-ms}~\kappa \label{kap:D2DTL}
\end{equation}
satisfies the bilinear equations (\ref{bl2:D2DTL}).
\end{theorem}
\begin{proof}
 From an appropriate Pl\"ucker relation, we can derive the bilinear
 equation with respect to the independent variables $(l_1,m,s)$ as
\begin{equation}
 \kappa^{m+1}\kappa_{l_1+1}-\kappa\kappa_{l_1+1}^{m+1}=b~\kappa^{s+1}\kappa_{l_1+1}^{m+1,s-1}.
\end{equation}
Since the linear relation (\ref{dSKP:linear1_molecule}) is essentially
 obtained by the change of variables $l_k\to -l_k$ ($k=2,3$) from (\ref{dSKP:linear1}), we have
\begin{equation}
(1+c_k(l_k)b)\kappa^{m+1}\kappa_{l_k+1} - \kappa\kappa_{l_k+1}^{m+1}
= c_k(l_k)b~\kappa^{s+1}\kappa_{l_k+1}^{m+1,s-1}\quad (k=2,3).
\end{equation}
Therefore $\tau$ defined by (\ref{kap:D2DTL}) satisfies (\ref{bl2:D2DTL}).
\end{proof}
\subsection{Reduction to Discrete Schwarzian KdV Equation}\label{sec:dSKP_red}
It is possible to obtain dSKdV by a certain reduction procedure from dSKP
(\ref{eqn:dSKP}). On the level of $z$, the reduction is achieved by applying three extra conditions,
which are obtained by imposing a certain symmetry in each of the independent variables $l_i$
($i=1,2,3$). The symmetry required is such that one of the independent variables $l_i$ ($i=1,2,3$)
of dSKP (\ref{eqn:dSKP}) can be negated in each of three conditions. In fact, only two conditions
are needed as the third is redundant, we only mention that there are three such conditions for
completeness.
\begin{proposition}
Let $z_{l_1,l_2,l_3}$ be a function satisfying (\ref{eqn:dSKP}). We
impose the following equations for $z$:
\begin{align}
 &\frac{(z_{l_1+1,l_2,l_3}-z_{l_1+1,l_2+1,l_3})(z_{l_1,l_2+1,l_3}-z_{l_1,l_2+1,l_3-1})(z_{l_1,l_2,l_3-1}-z_{l_1+1,l_2,l_3-1})}
{(z_{l_1+1,l_2+1,l_3}-z_{l_1,l_2+1,l_3})(z_{l_1,l_2+1,l_3-1}-z_{l_1,l_2,l_3-1})(z_{l_1+1,l_2,l_3-1}-z_{l_1+1,l_2,l_3})} =-1,\label{red1:dSKP}\\
 &\frac{(z_{l_1+1,l_2,l_3}-z_{l_1+1,l_2-1,l_3})(z_{l_1,l_2-1,l_3}-z_{l_1,l_2-1,l_3+1})(z_{l_1,l_2,l_3+1}-z_{l_1+1,l_2,l_3+1})}
{(z_{l_1+1,l_2-1,l_3}-z_{l_1,l_2-1,l_3})(z_{l_1,l_2-1,l_3+1}-z_{l_1,l_2,l_3+1})(z_{l_1+1,l_2,l_3+1}-z_{l_1+1,l_2,l_3})}=-1. \label{red1a:dSKP}
\end{align}
Then (\ref{eqn:dSKP}) is reduced to dSKdV (\ref{eqn:cross_ratio2}).
\end{proposition}
Note that a third condition, with $l_1\mapsto-l_1$, is consistent with (\ref{red1:dSKP}) and (\ref{red1a:dSKP}).
\begin{proof}
Let us fix $l_3$ and put $z_{l_1,l_2}=z_{l_1,l_2,l_3}$, $y_{l_1,l_2}=z_{l_1,l_2,l_3+1}$. We
drop the $l_3$ dependence in the expression of $y$ and $z$. Then (\ref{eqn:dSKP}),
(\ref{red1:dSKP})${}_{l_3+1}$ and (\ref{red1a:dSKP})${}_{l_2+1}$ can be written as
\begin{align}
&\frac{(z_{l_1+1}-z_{l_1+1,l_2+1})(z_{l_2+1}-y_{l_2+1})(y-y_{l_1+1})}
{(z_{l_1+1,l_2+1}-z_{l_2+1})(y_{l_2+1}-y)(y_{l_1+1}-z_{l_1+1})} =-1,\label{red2:dSKP}\\
 &\frac{(y_{l_1+1}-y_{l_1+1,l_2+1})(y_{l_2+1}-z_{l_2+1})(z-z_{l_1+1})}
{(y_{l_1+1,l_2+1}-y_{l_2+1})(z_{l_2+1}-z)(z_{l_1+1}-y_{l_1+1})} 
=-1,\label{red3:dSKP}\\
&\frac{(z_{l_1+1,l_2+1}-z_{l_1+1})(z-y)(y_{l_2+1}-y_{l_1+1,l_2+1})}{(z_{l_1+1}-z)(y-y_{l_2+1})(y_{l_1+1,l_2+1}-z_{l_1+1,l_2+1})}=-1,\label{red2a:dSKP}
\end{align}
respectively. One can eliminate $y$ from (\ref{red2:dSKP}) and (\ref{red3:dSKP}) as follows:
we first eliminate $y_{l_1+2,l_2+1}$ and $y_{l_1+1,l_2+2}$ from (\ref{red2:dSKP})$_{l_1+1,l_2+1}$ by using
(\ref{red3:dSKP})$_{l_1+1}$ and (\ref{red3:dSKP})$_{l_2+1}$. In the resulting equation, we use
(\ref{red2:dSKP})$_{l_1+1}$ and (\ref{red2:dSKP})$_{l_2+1}$ to eliminate $y_{l_1+2}$ and
 $y_{l_2+2}$. The resulting expression still contains $y_{l_1+1}$, $y_{l_2+1}$ and
 $y_{l_1+1,l_2+1}$, but they are eliminated by virtue of (\ref{red3:dSKP}), leaving the following
 expression in $z$ alone
\begin{equation}
\frac{(z_{l_1+2,l_2+2}-z_{l_1+2,l_2+1})(z_{l_1+1,l_2+2}-z_{l_2+2})(z_{l_1+2}-z_{l_1+1})(z_{l_2+1}-z)}
{(z_{l_1+2,l_2+2}-z_{l_1+1,l_2+2})(z_{l_1+2,l_2+1}-z_{l_1+2})(z_{l_2+2}-z_{l_2+1})(z_{l_1+1}-z)}=1. \label{red4:dSKP}
\end{equation}
Equation (\ref{red4:dSKP}) is rearranged in the following form
\begin{align}
& \frac{X_{l_1+1,l_2+1}}{X_{l_2+1}}=\frac{X_{l_1+1}}{X},\\
& X= \frac{(z-z_{l_1+1})(z_{l_1+1,l_2+1}-z_{l_2+1})}{(z_{l_1+1}-z_{l_1+1,l_2+1})(z_{l_2+1}-z)},
\end{align}
from which we obtain (\ref{eqn:cross_ratio2}).
\end{proof}
Note that the ansatz (\ref{dSKP_ansatz}) is reduced to (\ref{eqn:red_bilinear}) (or
(\ref{cross_ratio_ansatz2}) below) as follows.  Substituting (\ref{dSKP_ansatz}) into the
reduction condition (\ref{red1:dSKP}), we have
\begin{equation}
\frac{v_{l_2+1}}{w_{l_2+1}}~\frac{v_{l_1+1,l_3-1}}{w_{l_1+1,l_3-1}}= 
\frac{v_{l_1+1}}{w_{l_1+1}}~\frac{v_{l_2+1,l_3-1}}{w_{l_2+1,l_3-1}},
\end{equation}
which is solved by $w=R(l_1,l_2)\rho_3(l_3)v$, causing the ansatz (\ref{dSKP_ansatz}) to become
\begin{equation}
z_{l_i+1}-z=R(l_1,l_2)\rho_3(l_3)v_{l_i+1}v,\label{eqn:int_ansatz}
\end{equation}
where $R(l_1,l_2)$ and $\rho_3(l_3)$ are arbitrary functions. Now we apply the additional condition on $z$,
(\ref{red2a:dSKP}), into which we substitute (\ref{eqn:int_ansatz}). This produces a condition on
$R(l_1,l_2)$
\begin{equation}
\frac{R_{l_1+1,l_2+1}R}{R_{l_1+1}R_{l_2+1}}=1,\nonumber
\end{equation}
which implies that $R$ must be separable as $R(l_1,l_2)=\rho_1(l_1)\rho_2(l_2)$, where $\rho_1(l_1)$
and $\rho_2(l_2)$ are arbitrary.  Now introducing
$u=(\rho_1(l_1)\rho_2(l_2)\rho_3(l_3))^{\frac{1}{2}}v$, (\ref{eqn:int_ansatz}) can be rewritten
as
\begin{equation}
 z_{l_i+1}-z=\left(\frac{\rho_i(l_i)}{\rho_i(l_i+1)}\right)^{\frac{1}{2}}u_{l_i+1}u,
\end{equation}
which is equivalent to (\ref{eqn:red_bilinear}).

Let us consider the reduction on the level of the $\tau$ function and construct explicit solutions
to dSKdV. The above discussion and (\ref{dSKP:uv}) suggests that the reduction condition to be
imposed should be $\tau^{m+1}\Bumpeq \tau^{s+2}$, where $\Bumpeq$ means the equivalence up to 
gauge transformation. However, due to the difference of gauge invariance of the dSKP and dSKdV (and
their bilinear equations), the reduction cannot be applied in a straightforward manner. We apply the
reduction to the solutions in Theorem \ref{thm:Casorati_tau:dSKP} and \ref{thm:dSKP_molecule}
separately.

We first consider the Casorati determinant solution presented in
Theorem \ref{thm:Casorati_tau:dSKP}. Choose the entries of the determinant $\sigma$ as
\begin{equation}
\varphi_{i} = \alpha_i p_i^s \prod_{k=1}^3
  \prod_{j_k}^{l_k-1}(1-a_k(j_k)p_i)^{-1} \left(1+\frac{b}{p_i}\right)^{-m}
+\beta_i q_i^s \prod_{k=1}^3
  \prod_{j_k}^{l_k-1}(1-a_k(j_k)q_i)^{-1} \left(1+\frac{b}{q_i}\right)^{-m},
\end{equation}
where $p_i,q_i, \alpha_i,\beta_i$ ($i=1,\ldots,N$) are constants. We next impose the
condition
\begin{equation}
 \varphi_{i}^{m+1}\Bumpeq \varphi_i^{s+2},\label{reduction:Casorati_entry}
\end{equation}
which can be realized by choosing the parameters as
\begin{equation}
 q_i = -p_i - b.\label{reduction:parameters}
\end{equation}
Equation (\ref{reduction:Casorati_entry}) implies the following condition on $\sigma$:
\begin{equation}
 \sigma^{m+1}=C_N~\sigma^{s+2},\quad C_N=\prod_{i=1}^N p_i^{-2}\left(1+\frac{b}{p_i}\right)^{-1}.\label{reduction:Casorati}
\end{equation}
Using (\ref{reduction:Casorati}) to eliminate the $m$-dependence and neglecting $l_3$-dependence, equations (\ref{bl:sigma}) are reduced to
\begin{equation}
(1+a_k(l_k)b)\sigma_{l_k+1}^{s+2}\sigma - \sigma_{l_k+1}\sigma^{s+2} 
= a_k(l_k)b~\sigma_{l_k+1}^{s+1}\sigma^{s+1}\quad (k=1,2).\label{bl2:sigma}
\end{equation}
Introducing $z=z_{l_1,l_2}$ by
\begin{equation}
 z=\prod_{k=1}^2\prod_{j_k}^{l_k-1}(1+a_k(j_k)b)~\frac{\sigma^{s+2}}{\sigma},\label{z:cross_ratio1}
\end{equation}
then (\ref{bl2:sigma}) can be rewritten as
\begin{equation}
 z_{l_k+1}-z = \frac{a_k(l_k)b}{(1+a_k(l_k)b)^{\frac{1}{2}}}~u_{l_k+1}u\quad (k=1,2),\label{cross_ratio_ansatz2}
\end{equation}
where
\begin{equation}
u= \prod_{k=1}^2\prod_{j_k}^{l_k-1}(1+a_k(j_k)b)^{\frac{1}{2}}~\frac{\sigma^{s+1}}{\sigma}.
\end{equation}
From (\ref{cross_ratio_ansatz2}), we see that $z$ satisfies dSKdV (\ref{eqn:cross_ratio2})
with $\lambda(l_1)=a_1(l_1)^2/(1+a_1(l_1)b)$ and $\mu(l_2)=a_2(l_2)^2/(1+a_2(l_2)b)$. Summarizing the discussion, we have the
following theorem:
\begin{theorem}
 Let $\sigma=\sigma_{l_1,l_2}^s$ be an $N\times N$ determinant given by
\begin{equation}
 \sigma=\det (\varphi_{i,l_1,l_2}^{s+j-1})_{i,j=1,\ldots,N},
\end{equation}
\begin{equation}
\varphi_{i,l_1,l_2}^{s} = \alpha_i p_i^s \prod_{k=1}^2
  \prod_{j_k}^{l_k-1}(1-a_k(j_k)p_i)^{-1} +\beta_i q_i^s \prod_{k=1}^2  \prod_{j_k}^{l_k-1}(1-a_k(j_k)q_i)^{-1} ,
\end{equation}
where $p_i, q_i, \alpha_i,\beta_i$ are constants, $a_k(l_k)$ ($k=1,2$) are arbitrary functions in
the indicated variables and the parameters $p_i$ and $q_i$ are related as in
(\ref{reduction:parameters}). Then $\sigma$ satisfies the bilinear equations (\ref{bl2:sigma}), and
$z=z_{l_1,l_2}$ defined by (\ref{z:cross_ratio1}) satisfies dSKdV (\ref{eqn:cross_ratio2}) with
$\lambda(l_1)=a_1(l_1)^2/(1+a_1(l_1)b)$ and $\mu(l_2)=a_2(l_2)^2/(1+a_2(l_2)b)$.
\end{theorem}
We note that the soliton solutions to the autonomous case have been obtained in
\cite{HZ:ABS2,NAH:ABS1}. Similarly, one can construct the molecule type solution by applying the
reduction to the solution in Theorem \ref{thm:dSKP_molecule}.
\begin{theorem}
 Let $\kappa=\kappa_{l_1,l_2}^s$ be an $l_1\times l_1$ determinant given by
\begin{equation}
 \kappa=\det (\phi_{i,l_2}^{s+j-1})_{i,j=1,\ldots,l_1},
\end{equation}
\begin{align}
\phi_{i}^{s} =& \alpha_i p_i^s   \prod_{j}^{l_2-1}(1-c_2(j)p_i) 
+\beta_i q_i^s \prod_{j}^{l_2-1}(1-c_2(j)q_i) ,
\end{align}
where $p_i, q_i, \alpha_i,\beta_i$ are constants, $c_2(l_2)$ 
is an arbitrary function in $l_2$ and the parameters $p_i$ and $q_i$ are
 related as in (\ref{reduction:parameters}). Then $\kappa$ satisfies the bilinear equations 
\begin{align}
& c_1(l_1)\kappa^{s+2}\kappa_{l_1+1}-\kappa\kappa_{l_1+1}^{s+2}=b~\kappa^{s+1}\kappa_{l_1+1}^{s+1},\\
& (1+c_2(l_2)b)\kappa^{s+2}\kappa_{l_2+1} - \kappa\kappa_{l_2+1}^{s+2}= c_2(l_2)b~\kappa^{s+1}\kappa_{l_2+1}^{s+1},
\end{align}
where 
\begin{equation}
 c_1(l_1) = p_{l_1+1}^2\left(1+\frac{b}{p_{l_1+1}}\right).
\end{equation}
Moreover, 
\begin{equation}
  z_{l_1,l_2}=\prod_{j_1}^{l_1-1}c_1(j_1)^{-1}\prod_{j_2}^{l_2-1}(1+a_2(j_2)b)^{-1}~\frac{\kappa^{s+2}}{\kappa},
\end{equation}
satisfies dSKdV (\ref{eqn:cross_ratio2})
with $\lambda(l_1)=c_1(l_1)^{-1}$ and $\mu(l_2)=c_2(l_2)^2/(1+c_2(l_2))$.
\end{theorem}

The reduction process to dSKdV is somewhat delicate.  For the case of
dSKP, the coefficients of the bilinear equations can be removed by multiplying a certain gauge
factor to the $\tau$ function. For example, the bilinear equations (\ref{bl:sigma}) yield
(\ref{bl:D2DTL}) by using (\ref{sig:D2DTL}). For dSKdV, however, such a gauge
transformation does not work. For example, the bilinear equations (\ref{bl2:sigma}) can be rewritten
as
\begin{align}
\hat\tau_{l_k+1}^{s+2}\hat\tau - \hat\tau_{l_k+1}\hat\tau^{s+2} 
 = \frac{a_k(l_k)b}{(1+a_k(l_k)b)^{\frac{1}{2}}}~\hat\tau_{l_k+1}^{s+1}\hat\tau^{s+1}\ (k=1,2),\label{bl3:cross_ratio}
\end{align}
by introducing $\hat\tau=\hat\tau_{l_1,l_2}^{s}$ by
\begin{equation}
 \hat\tau = \prod_{k=1}^2\prod_{j_k}^{l_k-1}(1+a_k(j_k)b)^{\frac{s}{2}}~\sigma.
\end{equation}
Then we have
\begin{equation}
 z_{l_k+1}-z = \frac{a_k(l_k)b}{(1+a_k(l_k)b)^{\frac{1}{2}}}~u_{l_k+1}u,
\end{equation}
where
\begin{equation}
 z = \frac{\hat\tau^{s+2}}{\hat\tau},\quad u = \frac{\hat\tau^{s+1}}{\hat\tau}.
\end{equation}
The crucial difference from the case of dSKP is that the coefficient of the right hand side of
(\ref{bl3:cross_ratio}) cannot be removed by multiplying a gauge factor to $\tau$. Even for the
autonomous case, namely the case where $a_k(l_k)$ are constants, it is possible to remove the
coefficient of one of the two bilinear equations, but not possible to remove those of two equations
simultaneously.

%
Therefore, it is not appropriate to impose the condition $\tau^{m+1}\Bumpeq \tau^{s+2}$ on the
bilinear equations (\ref{bl:D2DTL}). Actually it is obvious that the bilinear equations
(\ref{bl3:cross_ratio}) cannot be obtained from naive reduction from (\ref{bl:D2DTL}).  We have
to apply the reduction to the Casorati determinant without its gauge factor instead of applying directly
to $\tau_{l_1,l_2,l_3}^{m,s}$ itself. Such inconsistencies involving gauge factors arising through the
reduction process can be seen for other non-autonomous discrete integrable
systems as well \cite{KO:non-autonomous_dKdV,KO:non-autonomous:RIMS}.
\begin{remark}
Konopelchenko and Schief discussed in \cite{Konopelchenko-Schief} the reduction from dSKP to the
dSKdV by imposing the condition
\begin{equation}
z_{l_2+1,l_3+1} = z, \label{z12}
\end{equation}
on (\ref{eqn:dSKP}). Using this condition to eliminate the $l_3$ dependence, (\ref{eqn:dSKP}) can be
rearranged in the form
\begin{equation}
 \frac{Y_{l_2-1}}{Y}=1,\quad Y = \frac{(z-z_{l_1+1})(z_{l_1+1,l_2+1}-z_{l_2+1})}{(z_{l_1+1}-z_{l_1+1,l_2+1})(z_{l_2+1}-z)},
\end{equation}
which yields the special case of (\ref{eqn:cross_ratio2})
\begin{equation}
 \frac{(z-z_{l_1+1})(z_{l_1+1,l_2+1}-z_{l_2+1})}{(z_{l_1+1}-z_{l_1+1,l_2+1})(z_{l_2+1}-z)}=\nu(l_1),
\end{equation}
where $\nu(l_1)$ is an arbitrary function. The condition (\ref{z12}) is the subcase of the
 condition (\ref{red1:dSKP}): it is easily verified that if $z$ satisfies (\ref{z12}) then
 (\ref{red1:dSKP}) is automatically satisfied. For the solution of dSKP given in Theorem
 \ref{thm:Casorati_tau:dSKP}, it can be shown that (\ref{z12}) is realized by taking
 $a_k(l_k)=a_k={\rm const.}$ ($k=2,3$), imposing (\ref{reduction:parameters}), and choosing $b$ as
 $b=-(\frac{1}{a_2}+\frac{1}{a_3})$.
\end{remark}
\section{Discrete Schwarzian KdV Equation in Painlev\'e Systems}
In this section, we consider the solutions of dSKdV which are expressed by $\tau$ functions of
certain Painlev\'e systems. We give two examples, one with the symmetry of the affine Weyl group of
type $(A_2+A_1)^{(1)}$, the other with that of type $D_4^{(1)}$, and construct explicit formulae of
solutions in terms of their $\tau$ functions. We note that these solutions are not directly related
to ones discussed in the previous section.
\subsection{Painlev\'e System of Type $(A_2+A_1)^{(1)}$}
The Painlev\'e system of type $(A_2+A_1)^{(1)}$
\cite{KNY:qp4,KK:qp3,K:qp3,KNT:projective_reduction,Sakai:Painleve} arises as a family of B\"acklund transformations
associated with a $q$-Painlev\'e III equation
\begin{equation}
\begin{split}
& g_{n+1}=\frac{q^{2N+1}c^2}{f_ng_n}\frac{1+a_0q^nf_n}{a_0q^n+f_n},\\ 
& f_{n+1}=\frac{q^{2N+1}c^2}{f_ng_{n+1}}\frac{1+a_0a_2q^{n-m}g_{n+1}}{a_0a_2q^{n-m}+g_{n+1}},
\end{split}\label{eqn:qp3}
\end{equation}
for the unknown functions $f_n=f_n(m,N)$, $g_n=g_n(m,N)$ and the independent variable
$n\in\mathbb{Z}$. Here, $m,N\in\mathbb{Z}$ and $a_0,a_2,c,q\in\mathbb{C}^{\times}$ are
parameters. The system of equations (\ref{eqn:qp3}) and its B\"acklund transformations can be
formulated as a birational representation of the extended affine Weyl group of type
$(A_2+A_1)^{(1)}$. We define the transformations $s_i$ ($i=0,1,2$) and $\pi$ on variables $f_j$
($j=0,1,2$) and parameters $a_k$ ($k=0,1,2$) by
\begin{equation}
\begin{array}{ll}
{\displaystyle s_i(a_j) = a_ja_i^{-a_{ij}}}, & {\displaystyle s_i(f_j)=f_j\left(\frac{a_i+f_j}{1+a_if_j}\right)^{u_{ij}},}\\[2mm]
{\displaystyle  \pi(a_j) = a_{j+1}}, & {\displaystyle \pi(f_j)=f_{j+1}, }
\end{array}
\end{equation}
for $i,j\in\mathbb{Z}/3\mathbb{Z}$. Here, $A=(a_{ij})_{i,j=0,1,2}$ and $U=(u_{ij})_{i,j=0,1,2}$ are
given by
\begin{align}
 A=\left(\begin{array}{ccc} 2& -1& -1\\-1 & 2 & -1\\ -1 & -1 & 2\end{array}\right),\\
 U=\left(\begin{array}{ccc} 0& 1& -1\\-1 & 0 & 1\\ 1 & -1 & 0\end{array}\right),
\end{align}
which are the Cartan matrix of type $A_2^{(1)}$ and the orientation matrix of the corresponding
Dynkin diagram, respectively. We also define the transformations $w_0$, $w_1$ and $r$ by
\begin{equation}
\begin{split}
& w_0(f_i) =
 \frac{a_ia_{i+1}(a_{i-1}a_i+a_{i-1}f_i+f_{i-1}f_i)}{f_{i-1}(a_ia_{i+1}+a_if_{i+1}+f_if_{i+1})},\\
& w_1(f_i) =  \frac{1+a_if_i+a_ia_{i+1}f_if_{i+1}}{a_ia_{i+1}f_{i+1}(1+a_{i-1}f_{i-1}+a_{i-1}a_if_{i-1}f_i)},\\
& r(f_i)=\frac{1}{f_i},\\
& w_0(a_i)=a_i,\quad w_1(a_i)=a_i,\quad r(a_i)=a_i,
\end{split}
\end{equation}
for $i\in\mathbb{Z}/3\mathbb{Z}$.
Then the group of birational transformations $\langle s_0,s_1,s_2,\pi,w_0,w_1,r\rangle$ form the
extended affine Weyl group $\widetilde{W}((A_2+A_1)^{(1)})$, namely the transformations satisfy the
fundamental relations
\begin{equation}
\begin{split}
&s_i^2=1,\ 
(s_is_{i+1})^3=1,\ 
\pi^3=1,\ 
\pi s_i=s_{i+1}\pi,\\
&w_0^2=w_1^2=r^2=1,\ rw_0=w_1r, 
\end{split}
\end{equation}
for $i\in\mathbb{Z}/3\mathbb{Z}$, and the actions of $\langle
s_0,s_1,s_2,\pi\rangle=\widetilde{W}(A_2^{(1)})$ and $\langle w_0,w_1,r\rangle=\widetilde{W}(A_1^{(1)})$
commute with each other. Note that 
\begin{equation}
 a_0a_1a_2=q,\quad f_0f_1f_2=qc^2
\end{equation}
are invariant with respect to the action of $\widetilde{W}\left((A_2+A_1)^{(1)}\right)$ and
$\widetilde{W}(A_2^{(1)})$, respectively. We define the translation operators $T_i$ ($i=1,2,3,4$) by
\begin{equation}
 T_1=\pi s_2s_1,\ T_2 = s_1\pi s_2,\ T_3=s_1s_2\pi,\ T_4=rw_0,
\end{equation}
whose actions on parameters $a_i$ $(i=0,1,2)$ and $c$ are given by
\begin{equation}
\begin{array}{l}
T_1:~(a_0,a_1,a_2,c)\mapsto(qa_0,q^{-1}a_1,a_2,c),\\
T_2:~(a_0,a_1,a_2,c)\mapsto(a_0,qa_1,q^{-1}a_2,c),\\
T_3:~(a_0,a_1,a_2,c)\mapsto(q^{-1}a_0,a_1,qa_2,c),\\
T_4:~(a_0,a_1,a_2,c)\mapsto(a_0,a_1,a_2,qc),
\end{array}
\end{equation}
respectively. Note that $T_i$ ($i=1,2,3,4$) commute with each other and $T_1T_2T_3=1$.
The action of $T_1$ on $f$-variables can be expressed as
\begin{equation}\label{eqn:qp32}
\begin{split}
&T_1(f_1)=\frac{qc^2}{f_1f_0}~\frac{1+a_0f_0}{a_0+f_0},\\
&T_1(f_0)=\frac{qc^2}{f_0T_1(f_1)}~\frac{1+a_2a_0T_1(f_1)}{a_2a_0+T_1(f_1)}. 
\end{split}
\end{equation}
Or, applying $T_1^nT_2^mT_4^N$ to (\ref{eqn:qp32}) and putting
\begin{equation}
T_1^nT_2^mT_4^N(f_i)=f_{i,n,m}^{N}\quad (i=0,1,2), 
\end{equation}
we obtain
\begin{equation}\label{eqn:qp33}
\begin{split}
&f_{1,n+1}=\frac{q^{2N+1}c^2}{f_{1}f_{0}}~
\frac{1+a_0q^nf_{0}}{a_0q^n+f_{0}},\\
&f_{0,n+1}=\frac{q^{2N+1}c^2}{f_{0}f_{1,n+1}}~
\frac{1+a_2a_0q^{n-m}f_{1,n+1}}{a_2a_0q^{n-m}+f_{1,n+1}}, 
\end{split}
\end{equation}
which is equivalent to $q$-P$_{\rm III}$ (\ref{eqn:qp3}). Here, we have employed the convention to
display only the shifted variables in $(n,m,N)$.

\begin{figure}[h]
\begin{center}
\includegraphics[width=0.3\textwidth]{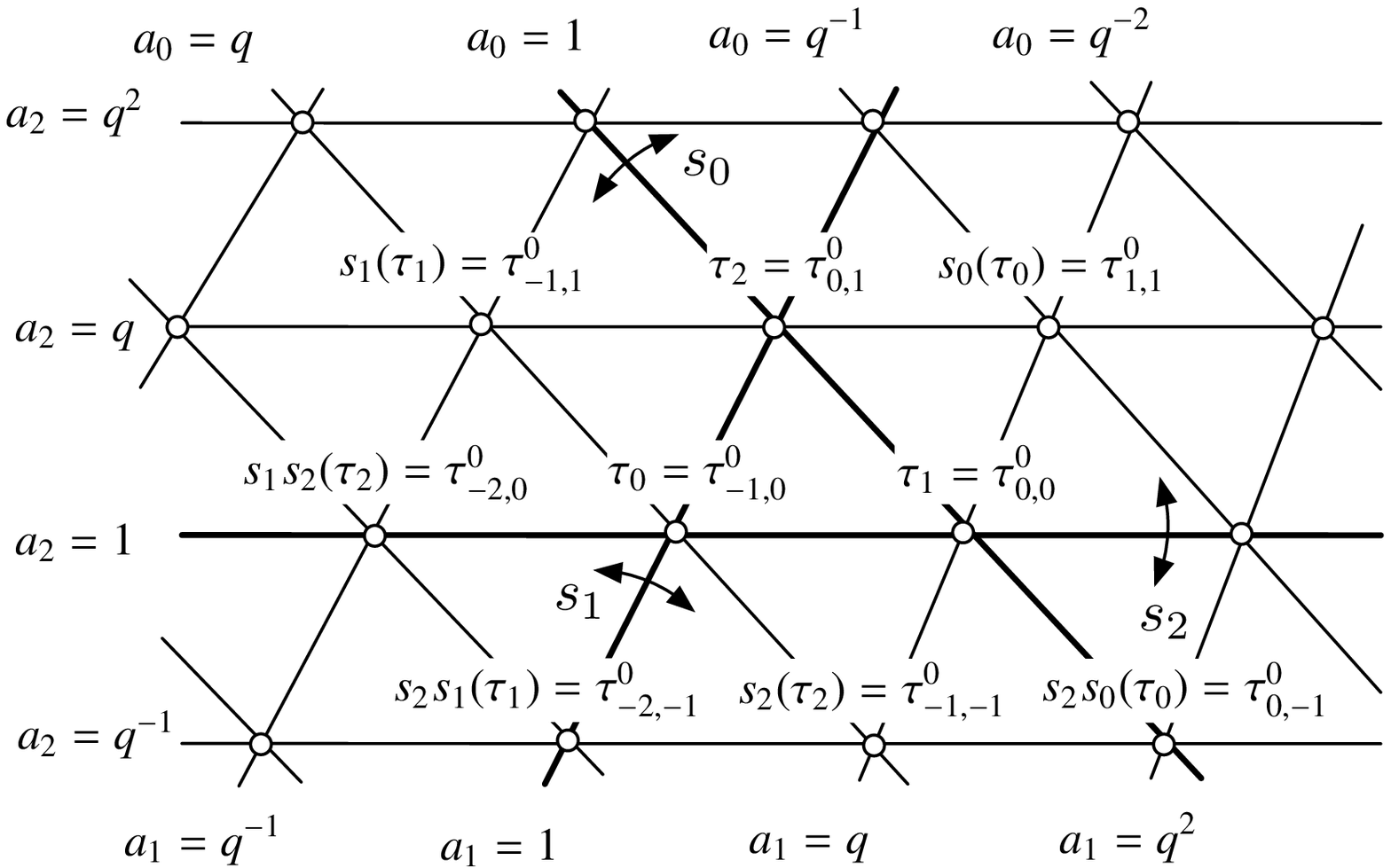} \quad
\raise40pt\hbox{\includegraphics[width=0.1\textwidth]{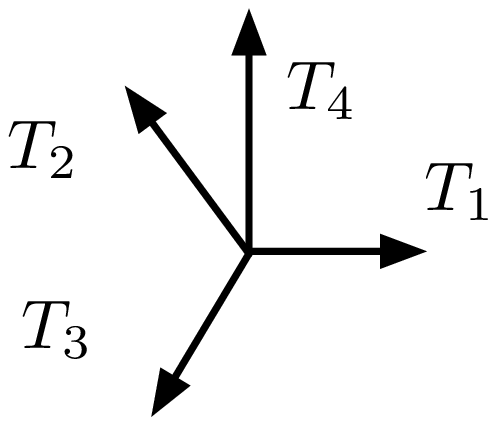}}
\end{center}
\caption{Configuration of the $\tau$ functions on the lattice with $N=0$.}\label{fig:tau_a2a1}
\end{figure}

It is possible to introduce the $\tau$ functions and lift the above representation of the affine
Weyl group on the level of the $\tau$ functions. We introduce the new variable $\tau_i$ and
$\overline{\tau}_i$ ($i\in\mathbb{Z}/3\mathbb{Z}$) with
\begin{equation}
 f_i = q^{\frac{1}{3}}c^{\frac{2}{3}}~\frac{\overline{\tau}_{i+1}\tau_{i-1}}{\tau_{i+1}\overline{\tau}_{i-1}}.
\end{equation}
Then the lift of the representation is realized by the following formulae:
\begin{equation}
 \begin{split}
  & s_i(\tau_i) =
  u_i^{\frac{1}{2}}\left(1+\frac{f_i}{a_i}\right)\frac{\tau_{i+1}\overline{\tau}_{i-1}}{\overline{\tau}_{i}},\\
  & s_i(\overline{\tau}_i) =
  v_i^{-\frac{1}{2}}\left(1+a_if_i\right)\frac{\tau_{i+1}\overline{\tau}_{i-1}}{\tau_{i}},\\
&s_i(\tau_j)=\tau_j,\quad s_i(\overline{\tau}_j)=\overline{\tau}_j\ (i\neq j),
 \end{split}
\end{equation}
\begin{equation}
 \begin{split}
  & w_0(\overline{\tau}_i)=\left(\frac{a_{i+1}}{a_{i-1}}\right)^{\frac{1}{3}}u_{i-1}
\left(1+\frac{f_{i-1}}{a_{i-1}}+\frac{f_{i-1}f_{i+1}}{a_{i-1}a_{i+1}}\right)\frac{\tau_i\tau_{i-1}}{\overline{\tau}_{i-1}},\\
  & w_1(\tau_i)=\left(\frac{a_{i+1}}{a_{i-1}}\right)^{\frac{1}{3}}v_{i+1}^{-1}
\left(1+a_{i-1}f_{i-1}+a_{i-1}a_{i+1}f_{i-1}f_{i+1}\right)\frac{\overline{\tau}_i\overline{\tau}_{i+1}}{\tau_{i+1}},\\
& w_0(\tau_i)=\tau_i,\  w_1(\overline{\tau}_i)=\overline{\tau}_i,\  r(\tau_i)=\tau_i,\  r(\overline{\tau}_i)=\overline{\tau_i},
 \end{split}
\end{equation}
where
\begin{equation}
 u_i = q^{-\frac{1}{3}}c^{-\frac{2}{3}}a_i,\quad v_i = q^{\frac{1}{3}}c^{\frac{2}{3}}a_i.
\end{equation}
We define the $\tau$ function on the lattice of type $A_2\times A_1$ by
\begin{equation}
T_1^nT_2^mT_4^N(\tau_1)=\tau_{n,m}^{N}.
\end{equation}
Note that $\tau_0=\tau_{-1,0}^{0}$,
$\tau_1=\tau_{0,0}^{0}$, $\tau_2=\tau_{0,1}^{0}$, $\overline{\tau}_0=\tau_{-1,0}^{1}$,
$\overline{\tau}_1=\tau_{0,0}^{1}$ and $\overline{\tau}_2=\tau_{0,1}^{1}$ (Fig.\ref{fig:tau_a2a1}).
From this construction, it follows that the $\tau$ functions satisfy various bilinear difference
equations, we refer to \cite{KNT:projective_reduction} for the list of these.

Now we show that dSKdV arises in the Painlev\'e system of type $(A_2+A_1)^{(1)}$:
\begin{theorem}\hfill
\begin{enumerate}
 \item The $\tau$ function satisfies the following bilinear equations:
\begin{align}
&\tau^{N+1}\tau_{n+1}^{N-1} - Q^{4N}\gamma^4\tau_{n+1}^{N+1}\tau^{N-1}
 + Q^{4n-2m-4N}\gamma^{-4}\alpha_0^6\alpha_1^{2}\alpha_2^4(Q^{12N}\gamma^{12}-1)\tau_{n+1}\tau=0,\label{bl1:a2a1}\\[2mm]
&\tau^{N+1}\tau_{m+1}^{N-1} - Q^{4N}\gamma^4\tau_{m+1}^{N+1}\tau^{N-1} + Q^{4m-2n-4N}
\gamma^{-4}\alpha_1^{2}\alpha_2^{-2}(Q^{12N}\gamma^{12}-1)\tau_{m+1}\tau=0.\label{bl2:a2a1}
\end{align}
Here, 
\begin{equation}
 \alpha_i = a_i^{\frac{1}{6}},\quad \gamma=c^{\frac{1}{6}},\quad Q=q^{\frac{1}{6}}.
\end{equation}
 \item We introduce $z=z_{n,m}$ by
\begin{equation}
 z = (Q^{4N+4}\gamma^4)^{n+m}~\frac{\tau^{N+2}}{\tau}.\label{z:a2a1}
\end{equation}
Then $z$ satisfies dSKdV
\begin{equation}
 \frac{(z-z_{n+1})(z_{n+1,m+1}-z_{m+1})}{(z_{n+1}-z_{n+1,m+1})(z_{m+1}-z)}=q^{2n-2m}a_0^2a_2^2.\label{eqn:cross_ratio_a2a1}
\end{equation}
\end{enumerate}
\end{theorem}
\begin{proof}
We take (B.17) and (B.20) of \cite{KNT:projective_reduction}:
\begin{align}
&\tau_{n+1}^{N+1}\tau^{N-1}
- Q^{m-2n-2N}\gamma^{-2}{\alpha_0}^{-3}{\alpha_1}^{-1}{\alpha_2}^{-2}\tau_{n+1,m+1}\tau_{m-1}
- Q^{4n-2m+4N}\gamma^{4}{\alpha_0}^{6}{\alpha_1}^{2}{\alpha_2}^{4}\tau_{n+1}\tau = 0, \\
&\tau^{N+1}\tau_{n+1}^{N-1}
- Q^{-2n+m+2N}\gamma^{2}{\alpha_0}^{-3}{\alpha_1}^{-1}{\alpha_2}^{-2}\tau_{m-1}\tau_{n+1,m+1}
 - Q^{4n-2m-4N}\gamma^{-4}{\alpha_0}^{6}{\alpha_1}^{2}{\alpha_2}^{4}\tau\tau_{n+1} = 0.
\end{align}
We obtain (\ref{bl1:a2a1}) by eliminating $\tau_{n+1,m+1}\tau_{m-1}$ from the above
 equations. Similarly, (\ref{bl2:a2a1}) can be derived by taking (B.18) and (B.21) of
 \cite{KNT:projective_reduction},
\begin{align}
  &\tau_{m+1}^{N+1}\tau^{N-1}
-  Q^{n-2m-2N}\gamma^{-2}{\alpha_1}^{-1}\alpha_2\tau_{n-1}\tau_{n+1,m+1}
- Q^{-2n+4m+4N}\gamma^{4}{\alpha_1}^{2}{\alpha_2}^{-2}\tau_{m+1}\tau = 0,\\
&\tau^{N+1}\tau_{m+1}^{N-1}
- Q^{n-2m+2N-1}\gamma^{2}\alpha_0{\alpha_2}^{2}\tau_{n+1,m+1}\tau_{n-1} 
-  Q^{-2n+4m-4N+2}\gamma^{-4}{\alpha_0}^{-2}{\alpha_2}^{-4}\tau\tau_{m+1} = 0,
\end{align}
and eliminating $\tau_{n+1,m+1}\tau_{n-1}$. Dividing (\ref{bl1:a2a1})
 and (\ref{bl2:a2a1}) by $\tau_{n+1}^{N-1}\tau^{N-1}$ and $\tau_{m+1}^{N-1}\tau^{N-1}$,
 respectively, shifting $N\to N+1$, and using (\ref{z:a2a1}), we have
\begin{align}
z - z_{n+1} =& -(Q^{4N+4}\gamma^4)^{n+m} Q^{4n-2m-4N-4} 
\gamma^{-4}\alpha_0^6\alpha_1^{2}\alpha_2^4(Q^{12N+12}\gamma^{12}-1) u_{n+1}u,\label{zu1:a2a1}\\
z - z_{m+1} =& -(Q^{4N+4}\gamma^4)^{n+m} Q^{4m-2n-4N-4} 
\gamma^{-4}\alpha_1^{2}\alpha_2^{-2}(Q^{12N+12}\gamma^{12}-1) u_{m+1}u,\label{zu2:a2a1}
\end{align}
where 
\begin{equation}
 u=\frac{\tau^{N+1}}{\tau}.
\end{equation}
Then it is easy to verify from (\ref{zu1:a2a1}) and (\ref{zu2:a2a1}) that $z$ actually satisfies
(\ref{eqn:cross_ratio_a2a1}).
\end{proof}

\subsection{Painlev\'e System of Type $D_4^{(1)}$}
The relationship between dSKdV
\begin{equation}
\dfrac{(z_{n,m}-z_{n+1,m})(z_{n+1,m+1}-z_{n,m+1})}
      {(z_{n+1,m}-z_{n+1,m+1})(z_{n,m+1}-z_{n,m})}
=\dfrac{1}{t},
\label{LS-KdV}
\end{equation}
and the sixth Painlev\'e equation (P$_{\rm VI}$)
\begin{equation}
\begin{array}{l}
\dfrac{d^2q}{dt^2}=
\dfrac{1}{2}\left(\dfrac{1}{q}+\dfrac{1}{q-1}+\dfrac{1}{q-t}\right)\!\!
\left(\dfrac{dq}{dt}\right)^2\\[4mm]
~
-\left(\dfrac{1}{t}+\dfrac{1}{t-1}+\dfrac{1}{q-t}\right)\!\dfrac{dq}{dt}
+\dfrac{q(q-1)(q-t)}{2t^2(t-1)^2}\\[4mm]
~\times
\left[\kappa_{\infty}^2-\kappa_0^2\dfrac{t}{q^2}
     +\kappa_1^2\dfrac{t-1}{(q-1)^2}+(1-\theta^2)\dfrac{t(t-1)}{(q-t)^2}\right]
\end{array}   \label{P6}
\end{equation}
is discussed in \cite{NRGO:P6}.  In a word, dSKdV is a part of the B\"acklund transformations of
P$_{\rm VI}$, which is formulated as a birational representation of the extended affine Weyl group of
type $D_4^{(1)}$. In this subsection, we construct a class of the particular solutions to dSKdV
in terms of the $\tau$ functions of P$_{\rm VI}$.

As a preparation, we give a brief review of the B\"acklund transformations and some of the bilinear
equations for the $\tau$ functions \cite{Masuda}. It is well-known that P$_{\rm VI}$ (\ref{P6})
is equivalent to the Hamilton system
\begin{equation}
q'=\dfrac{\partial H}{\partial p},\quad p'=-\dfrac{\partial H}{\partial q},
\quad '=t(t-1)\dfrac{d}{dt},
\end{equation}
whose Hamiltonian is given by 
\begin{equation}
H=f_0f_3f_4f_2^2-[\alpha_4f_0f_3+\alpha_3f_0f_4+(\alpha_0-1)f_3f_4]f_2+\alpha_2(\alpha_1+\alpha_2)f_0.
\end{equation}
Here $f_i$ and $\alpha_i$ are defined by 
\begin{equation}
f_0=q-t,\quad f_3=q-1,\quad f_4=q,\quad f_2=p,
\end{equation}
and
\begin{equation}
\alpha_0=\theta,\quad\alpha_1=\kappa_{\infty},\quad
\alpha_3=\kappa_1,\quad\alpha_4=\kappa_0
\end{equation}
with  $\alpha_0+\alpha_1+2\alpha_2+\alpha_3+\alpha_4=1$. The B\"acklund transformations of P$_{\rm VI}$ are described by 
\begin{equation}
s_i(\alpha_j)=\alpha_j-a_{ij}\alpha_i\quad (i,j=0,1,2,3,4),
\end{equation}
\begin{equation}
s_2(f_i)=f_i+\dfrac{\alpha_2}{f_2},\quad
s_i(f_2)=f_2-\dfrac{\alpha_i}{f_i}~~(i=0,3,4),
\end{equation}
\begin{equation}
\begin{array}{ll}
s_5:\!\!\!\!
&\alpha_0\leftrightarrow\alpha_1,\quad\alpha_3\leftrightarrow\alpha_4,\\[2mm]
&f_2\mapsto -\dfrac{f_0(f_2f_0+\alpha_2)}{t(t-1)},\quad
f_4\mapsto t\dfrac{f_3}{f_0},\\[3mm]
s_6:\!\!\!\!
&\alpha_0\leftrightarrow\alpha_3,\quad\alpha_1\leftrightarrow\alpha_4,\\[2mm]
&f_2\mapsto -\dfrac{f_4(f_4f_2+\alpha_2)}{t},\quad 
f_4\mapsto\dfrac{t}{f_4},\\[3mm]
s_7:\!\!\!\!
&\alpha_0\leftrightarrow\alpha_4,\quad\alpha_1\leftrightarrow\alpha_3,\\[2mm]
&f_2\mapsto\dfrac{f_3(f_3f_2+\alpha_2)}{t-1},\quad 
f_4\mapsto\dfrac{f_0}{f_3},
\end{array}
\end{equation}
where $A=(a_{ij})_{i,j=0}^4$ is the Cartan matrix of type $D^{(1)}_4$. Then the group of birational transformations $\langle s_0,\ldots,s_7\rangle$ generate the extended affine Weyl group $\widetilde{W}(D^{(1)}_4)$. In fact, these generators satisfy the fundamental relations 
\begin{equation}
\begin{array}{l}
s_i^2=1\quad(i=0,\ldots,7),\\
s_is_2s_i=s_2s_is_2\quad(i=0,1,3,4),
\end{array}
\end{equation}
and 
\begin{equation}
\begin{array}{l}
s_5s_{\{0,1,2,3,4\}}=s_{\{1,0,2,4,3\}}s_5,\\
s_6s_{\{0,1,2,3,4\}}=s_{\{3,4,2,0,1\}}s_6,\\
s_7s_{\{0,1,2,3,4\}}=s_{\{4,3,2,1,0\}}s_7,\\[1mm]
s_5s_6=s_6s_5,\quad s_5s_7=s_7s_5,\quad s_6s_7=s_7s_6.
\end{array}
\end{equation}
Let us introduce the variables $\tau_i\,(i=0,1,2,3,4)$ via the Hamiltonian, so that the action of $\widetilde{W}(D^{(1)}_4)$ is given by 
\begin{equation}
\begin{array}{l}
s_0(\tau_0)=f_0\dfrac{\tau_2}{\tau_0},\quad
s_1(\tau_1)=\dfrac{\tau_2}{\tau_1},\\[3mm]
s_2(\tau_2)=\dfrac{f_2}{\sqrt{t}}
\dfrac{\tau_0\tau_1\tau_3\tau_4}{\tau_2},\\[3mm]
s_3(\tau_3)=f_3\dfrac{\tau_2}{\tau_3},\quad
s_4(\tau_4)=f_4\dfrac{\tau_2}{\tau_4},
\end{array}
\end{equation}
and 
\begin{equation}
\begin{array}{ll}
s_5:\!\!\!\!
&\tau_0\mapsto [t(t-1)]^{\frac{1}{4}}\tau_1,\quad
\tau_1\mapsto [t(t-1)]^{-\frac{1}{4}}\tau_0,\\[1mm]
&\tau_3\mapsto t^{-\frac{1}{4}}(t-1)^{\frac{1}{4}}\tau_4,\quad
\tau_4\mapsto t^{\frac{1}{4}}(t-1)^{-\frac{1}{4}}\tau_3,\\[1mm]
&\tau_2\mapsto [t(t-1)]^{-\frac{1}{2}}f_0\tau_2,
\end{array}
\end{equation}
\begin{equation}
\begin{array}{ll}
s_6:\!\!\!\!
&\tau_0\mapsto it^{\frac{1}{4}}\tau_3,\quad
\tau_3\mapsto -it^{-\frac{1}{4}}\tau_0,\\[1mm]
&\tau_1\mapsto t^{-\frac{1}{4}}\tau_4,\quad
\tau_4\mapsto t^{\frac{1}{4}}\tau_1,\quad
\tau_2\mapsto t^{-\frac{1}{2}}f_4\tau_2,
\end{array}
\end{equation}
\begin{equation}
\begin{array}{ll}
s_7:\!\!\!\!
&\tau_0\mapsto (-1)^{-\frac{3}{4}}(t-1)^{\frac{1}{4}}\tau_4,\quad
\tau_4\mapsto(-1)^{\frac{3}{4}}(t-1)^{-\frac{1}{4}}\tau_0,\\[1mm]
&\tau_1\mapsto(-1)^{\frac{3}{4}}(t-1)^{-\frac{1}{4}}\tau_3,\quad
\tau_3\mapsto(-1)^{-\frac{3}{4}}(t-1)^{\frac{1}{4}}\tau_1,\\[1mm]
&\tau_2\mapsto -i(t-1)^{-\frac{1}{2}}f_3\tau_2. 
\end{array}
\end{equation}
We note that some of the fundamental relations are modified 
\begin{equation}
s_is_2(\tau_2)=-s_2s_i(\tau_2)\quad(i=5,6,7),
\end{equation}
and
\begin{equation}
\begin{array}{l}
s_5s_6\tau_{\{0,1,2,3,4\}}=\{i,-i,-1,-i,i\}s_6s_5\tau_{\{0,1,2,3,4\}},\\[1mm]
s_5s_7\tau_{\{0,1,2,3,4\}}=\{i,-i,-1,i,-i\}s_7s_5\tau_{\{0,1,2,3,4\}},\\[1mm]
s_6s_7\tau_{\{0,1,2,3,4\}}=\{-i,-i,-1,i,i\}s_7s_6\tau_{\{0,1,2,3,4\}}.
\end{array}
\end{equation}
From the above formulation, one can obtain the bilinear equations for the $\tau$ functions. As examples, we have 
\begin{equation}
\begin{array}{l}
\alpha_2t^{-\frac{1}{2}}\tau_3\tau_4
=s_1(\tau_1)s_2s_0(\tau_0)-s_0(\tau_0)s_2s_1(\tau_1),\\[1mm]
\alpha_2t^{\frac{1}{2}}\tau_1\tau_3
=s_4(\tau_4)s_2s_0(\tau_0)-s_0(\tau_0)s_2s_4(\tau_4),\\[1mm]
\alpha_2t^{-\frac{1}{2}}\tau_0\tau_1
=s_4(\tau_4)s_2s_3(\tau_3)-s_3(\tau_3)s_2s_4(\tau_4),\\[1mm]
\alpha_2t^{-\frac{1}{2}}\tau_0\tau_4
=s_1(\tau_1)s_2s_3(\tau_3)-s_3(\tau_3)s_2s_1(\tau_1).
\end{array}   \label{bi:BT:a2}
\end{equation}

Let us introduce the translation operators
\begin{equation}
\begin{array}{c}
\widehat{T}_{13}=s_1s_2s_0s_4s_2s_1s_7,\quad 
\widehat{T}_{40}=s_4s_2s_1s_3s_2s_4s_7,\\
\widehat{T}_{34}=s_3s_2s_0s_1s_2s_3s_5,\quad 
T_{14}=s_1s_4s_2s_0s_3s_2s_6,
\end{array}   \label{translation}
\end{equation}
whose action on the parameters $\vec{\alpha}=(\alpha_0,\alpha_1,\alpha_2,\alpha_3,\alpha_4)$ is given by 
\begin{equation}
\begin{array}{c}
\widehat{T}_{13}(\vec{\alpha})=\vec{\alpha}+(0,1,0,-1,0),\\[1mm]
\widehat{T}_{40}(\vec{\alpha})=\vec{\alpha}+(-1,0,0,0,1),\\[1mm]
\widehat{T}_{34}(\vec{\alpha})=\vec{\alpha}+(0,0,0,1,-1),\\[1mm]
T_{14}(\vec{\alpha})=\vec{\alpha}+(0,1,-1,0,1).
\end{array}
\end{equation}
We denote $\tau_{k,l,m,n}=T_{14}^n\widehat{T}_{34}^m\widehat{T}_{40}^l\widehat{T}_{13}^k(\tau_0)\,(k,l,m,n\in\Z)$. 

\begin{theorem}
Let $z_{n,m}$ be
\begin{equation}
z_{n,m}=\left\{
\begin{array}{l}
(-1)^{n-1}
\dfrac{\tau_{-\frac{n+m}{2}-1,-\frac{n+m}{2}-1,-2,n}}
      {\tau_{-\frac{n+m}{2},-\frac{n+m}{2},0,n}}\quad \mbox{($n+m$ is even)},\\[4mm]
(-1)^{n-1}
\dfrac{\tau_{-\frac{n+m+1}{2},-\frac{n+m+1}{2}-1,-2,n}}
      {\tau_{-\frac{n+m-1}{2},-\frac{n+m+1}{2},0,n}}\quad 
\mbox{($n+m$ is odd)},
\end{array}
\right..
\end{equation}
Then, $z_{n,m}$ satisfies the dSKdV (\ref{LS-KdV}). 
\end{theorem}

\begin{proof}
Note that we get the bilinear equations 
\begin{equation}
\begin{array}{l}
(\alpha_0+\alpha_2+\alpha_4)\,t^{-\frac{1}{2}}\tau_3\tau_{4,4}
=\tau_{1,1}\tau_{420,0}-\tau_0\tau_{0421,1},\\[1mm]
(\alpha_0+\alpha_2+\alpha_4)\,t^{\frac{1}{2}}\tau_1\tau_3
=\tau_4\tau_{420,0}-\tau_0\tau_{024,4},
\end{array}   \label{bi_1}
\end{equation}
and 
\begin{equation}
\begin{array}{l}
(\alpha_0+\alpha_2+\alpha_4)t^{-\frac{1}{2}}\tau_{0,0}\tau_1 =\tau_4\tau_{0423,3}-\tau_{3,3}\tau_{024,4},\\[1mm]
(\alpha_0+\alpha_2+\alpha_4)t^{-\frac{1}{2}}\tau_{0,0}\tau_{4,4}=\tau_{1,1}\tau_{0423,3}-\tau_{3,3}\tau_{0421,1},
\end{array}   \label{bi_2}
\end{equation}
by applying the transformation $s_0s_4$ to (\ref{bi:BT:a2}), where we denote $s_j\cdots s_i(\tau_i)$
 by $\tau_{j\cdots i,i}$.

First, we consider the case where $n+m$ is even. The bilinear equations (\ref{bi_1}) can be expressed as 
\begin{equation}
\begin{array}{l}
\dfrac{\tau_{-1,-1,-2,0}}{\tau_{0,0,0,0}}+\dfrac{\tau_{-1,-2,-2,1}}{\tau_{0,-1,0,1}}=-(\alpha_0+\alpha_2+\alpha_4)\,t^{-\frac{1}{2}}
\dfrac{\tau_{0,-1,-1,0}\tau_{-1,-1,-1,1}}{\tau_{0,0,0,0}\tau_{0,-1,0,1}},\\[4mm]
\dfrac{\tau_{-1,-2,-2,0}}{\tau_{0,-1,0,0}}
-\dfrac{\tau_{-1,-1,-2,0}}{\tau_{0,0,0,0}}=(\alpha_0+\alpha_2+\alpha_4)\dfrac{\tau_{-1,-1,-1,0}\tau_{0,-1,-1,0}}{\tau_{0,0,0,0}\tau_{0,-1,0,0}}.
\end{array}   \label{bi_1'}
\end{equation}
Apply the translation $T_{14}^n\widehat{T}_{40}^{-N}\widehat{T}_{13}^{-N}$ to the above equations and put $N=\dfrac{n+m}{2}$. Then we get 
\begin{equation}
\begin{array}{l}
z_{n,m}-z_{n+1,m}
=(-1)^n(\alpha_0+\alpha_2+\alpha_4)\,t^{-\frac{1}{2}}\dfrac{\tau_{-N,-N-1,-1,n}\tau_{-N-1,-N-1,-1,n+1}}
{\tau_{-N,-N,0,n}\tau_{-N,-N-1,0,n+1}},\\[4mm]
z_{n,m+1}-z_{n,m}=(-1)^{n-1}(\alpha_0+\alpha_2+\alpha_4)\dfrac{\tau_{-N-1,-N-1,-1,n}\tau_{-N,-N-1,-1,n}}{\tau_{-N,-N,0,n}\tau_{-N,-N-1,0,n}}.
\end{array}
\end{equation}
Similarly, the bilinear equations (\ref{bi_2}) are expressed as 
\begin{equation}
\begin{array}{l}
\dfrac{\tau_{-2,-2,-2,1}}{\tau_{-1,-1,0,1}}
+\dfrac{\tau_{-1,-2,-2,0}}{\tau_{0,-1,0,0}}
=-(\alpha_0+\alpha_2+\alpha_4)t^{-\frac{1}{2}}
\dfrac{\tau_{-1,-1,-1,0}\tau_{-1,-2,-1,1}}{\tau_{0,-1,0,0}\tau_{-1,-1,0,1}},\\[4mm]
\dfrac{\tau_{-1,-2,-2,1}}{\tau_{0,-1,0,1}}
-\dfrac{\tau_{-2,-2,-2,1}}{\tau_{-1,-1,0,1}}=(\alpha_0+\alpha_2+\alpha_4)
\dfrac{\tau_{-1,-2,-1,1}\tau_{-1,-1,-1,1}}{\tau_{0,-1,0,1}\tau_{-1,-1,0,1}}.
\end{array}   \label{bi_2'}
\end{equation}
Then we get 
\begin{equation}
\begin{array}{l}
z_{n+1,m+1}-z_{n,m+1}
=(-1)^{n-1}(\alpha_0+\alpha_2+\alpha_4)t^{-\frac{1}{2}}\dfrac{\tau_{-N-1,-N-1,-1,n}\tau_{-N-1,-N-2,-1,n+1}}
      {\tau_{-N,-N-1,0,n}\tau_{-N-1,-N-1,0,n+1}},\\[4mm]
z_{n+1,m}-z_{n+1,m+1}
=(-1)^n(\alpha_0+\alpha_2+\alpha_4)
\dfrac{\tau_{-N-1,-N-2,-1,n+1}\tau_{-N-1,-N-1,-1,n+1}}
      {\tau_{-N,-N-1,0,n+1}\tau_{-N-1,-N-1,0,n+1}}.
\end{array}
\end{equation}
Thus we find that $z_{n,m}$ satisfies 
dSKdV (\ref{LS-KdV}) when $n+m$ is even. 

Next, we consider the case where $n+m$ is odd. From the bilinear equations (\ref{bi_2'}), we get 
\begin{equation}
\begin{array}{l}
z_{n,m}-z_{n+1,m}
=(-1)^n(\alpha_0+\alpha_2+\alpha_4)t^{-\frac{1}{2}}\dfrac{\tau_{-N,-N,-1,n}\tau_{-N,-N-1,-1,n+1}}
      {\tau_{-N+1,-N,0,n}\tau_{-N,-N,0,n+1}},\\[4mm]
z_{n,m+1}-z_{n,m}
=(-1)^n(\alpha_0+\alpha_2+\alpha_4)
\dfrac{\tau_{-N,-N-1,-1,n}\tau_{-N,-N,-1,n}}
      {\tau_{-N+1,-N,0,n}\tau_{-N,-N,0,n}},
\end{array}
\end{equation}
where we denote $N=\dfrac{n+m+1}{2}$. We also have 
\begin{equation}
\begin{array}{l}
z_{n+1,m+1}-z_{n,m+1}
=(-1)^{n-1}(\alpha_0+\alpha_2+\alpha_4)\,t^{-\frac{1}{2}}
\dfrac{\tau_{-N,-N-1,-1,n}\tau_{-N-1,-N-1,-1,n+1}}
      {\tau_{-N,-N,0,n}\tau_{-N,-N-1,0,n+1}},\\[4mm]
z_{n+1,m}-z_{n+1,m+1}
=(-1)^{n-1}(\alpha_0+\alpha_2+\alpha_4)
\dfrac{\tau_{-N-1,-N-1,-1,n+1}\tau_{-N,-N-1,-1,n+1}}
      {\tau_{-N,-N,0,n+1}\tau_{-N,-N-1,0,n+1}},
\end{array}
\end{equation}
from the bilinear equations (\ref{bi_1'}). 
Thus, we find that $z_{n,m}$ satisfies dSKdV (\ref{LS-KdV}). 
\end{proof}

By a similar argument, we obtain the following Theorem. 
\begin{theorem}
Let $z_{n,m}$ be
\begin{equation}
z_{n,m}=\left\{
\begin{array}{ll}
(-1)^{\binom{n-m+2}{2}}
\dfrac{\tau_{-\frac{n+m}{2}+1,-\frac{n+m}{2}-1,0,m}}
      {\tau_{-\frac{n+m}{2},-\frac{n+m}{2},0,m}}\quad
\mbox{($n+m$ is even)},\\[4mm]
(-1)^{\binom{n-m+2}{2}}
\dfrac{\tau_{-\frac{n+m-1}{2},-\frac{n+m+1}{2}-1,-1,m}}
      {\tau_{-\frac{n+m+1}{2},-\frac{n+m+1}{2},-1,m}}\quad
\mbox{($n+m$ is odd)},
\end{array}
\right..
\end{equation}
Then, $z_{n,m}$ satisfies dSKdV (\ref{LS-KdV}). 
\end{theorem}

\begin{proof}
By applying $s_0s_1$ to (\ref{bi:BT:a2}), we get 
\begin{equation}
\begin{array}{l}
\dfrac{\tau_{1,-1,0,0}}{\tau_{0,0,0,0}}
-\dfrac{\tau_{0,-2,-1,0}}{\tau_{-1,-1,-1,0}}
=(-1)^{\frac{1}{2}}(\alpha_0+\alpha_1+\alpha_2)\,t^{-\frac{1}{2}}
\dfrac{\tau_{0,-1,-1,0}\tau_{0,-1,0,0}}
      {\tau_{0,0,0,0}\tau_{-1,-1,-1,0}},\\[4mm]
\dfrac{\tau_{1,-1,0,0}}{\tau_{0,0,0,0}}
+\dfrac{\tau_{0,-2,-1,1}}{\tau_{-1,-1,-1,1}}
=(-1)^{\frac{1}{2}}(\alpha_0+\alpha_1+\alpha_2)
\dfrac{\tau_{0,-1,-1,0}\tau_{0,-1,0,1}}{\tau_{0,0,0,0}\tau_{-1,-1,-1,1}},
\end{array}
\end{equation}
\begin{equation}
\begin{array}{l}
\dfrac{\tau_{0,-2,0,1}}{\tau_{-1,-1,0,1}}
+\dfrac{\tau_{0,-2,-1,1}}{\tau_{-1,-1,-1,1}}
=(-1)^{\frac{3}{2}}(\alpha_0+\alpha_1+\alpha_2)t^{-\frac{1}{2}}
\dfrac{\tau_{-1,-2,-1,1}\tau_{0,-1,0,1}}
      {\tau_{-1,-1,0,1}\tau_{-1,-1,-1,1}},\\[4mm]
\dfrac{\tau_{0,-2,-1,0}}{\tau_{-1,-1,-1,0}}
-\dfrac{\tau_{0,-2,0,1}}{\tau_{-1,-1,0,1}}
=(-1)^{\frac{1}{2}}(\alpha_0+\alpha_1+\alpha_2)
\dfrac{\tau_{0,-1,0,0}\tau_{-1,-2,-1,1}}{\tau_{-1,-1,-1,0}\tau_{-1,-1,0,1}},
\end{array}
\end{equation}
which leads us to the above theorem. 
\end{proof}
\section*{Acknowledgements}
This work is partially supported by JSPS Grant-in-Aid for Scientifuc Research No. 19340039 and
21656027. M. Hay appreciatively acknowledges support from the Global COE Program ``Education and
Research Hub for Mathematics-for-Industry'' from the Ministry of Education, Culture, Sports, Science
and Technology, Japan.


\end{document}